\documentclass[twocolumn,prl]{revtex4}

\usepackage{graphicx}
\usepackage{dcolumn}
\usepackage{bm}
\usepackage{multirow}
\usepackage{color}
\usepackage[normalem]{ulem}
\usepackage{amsmath}
\usepackage{gensymb}
\usepackage[colorlinks=true]{hyperref}

\begin{document}

\title{Second harmonic generation control in twisted bilayers\\ of transition metal dichalcogenides}
\author{Ioannis Paradisanos$^{1}$}
\email{paradeis@insa-toulouse.fr}
\author{Andres Manuel  Saiz Raven$^{1}$}
\author{Thierry Amand$^{1}$}
\author{Cedric Robert$^1$}
\author{Pierre Renucci$^{1}$}
\author{Kenji Watanabe$^2$}
\author{Takashi Taniguchi$^3$}
\author{Iann C. Gerber$^1$}
\author{Xavier Marie$^1$}
\author{Bernhard Urbaszek$^1$}
\email{urbaszek@insa-toulouse.fr}

\affiliation{%
$^1$Universit\'e de Toulouse, INSA-CNRS-UPS, LPCNO, 135 Avenue Rangueil, 31077 Toulouse, France}
\affiliation{%
$^2$Research Center for Functional Materials, National Institute for Materials Science, 1-1 Namiki, Tsukuba 305-0044, Japan}
\affiliation{%
$^3$International Center for Materials Nanoarchitectonics, National Institute for Materials Science, 1-1 Namiki, Tsukuba 305-0044, Japan}

\begin{abstract}
The twist angle in transition metal dichalcogenide (TMD) heterobilayers is a compelling degree of freedom that determines electron correlations and the period of lateral confinement of moiré excitons. 
Here we perform polarization-resolved second harmonic generation (SHG) spectroscopy of MoS$_{2}$/WSe$_{2}$ heterostructures. 
We demonstrate that by choosing suitable laser energies the twist angle between two monolayers can be measured directly  on the assembled heterostructure. We show that the amplitude and polarization of the SHG signal from the heterostructure are determined by the twist angle between the layers and exciton resonances at the SH energy. 
For heterostructures with close to zero twist angle, we observe changes of exciton resonance energies and the appearance of new resonances in the linear and non-linear susceptibilities.  
\end{abstract}

\maketitle

\textbf{INTRODUCTION} \\
The optical and electronic properties of van der Waals (vdW) materials can be tuned by adjusting the twist angle for bilayer heterostructures \cite{tran2020moire,shree2020guide}.  A moiré superlattice forms by stacking two transition metal dichalcogenide (TMD) monolayers with a finite twist angle and/or different lattice constants \cite{oster1963moire}. This periodic pattern can host new quantum phenomena in two-dimensional heterostructures \cite{yu2017moire} and homobilayers. Correlated states and superconductivity are investigated in twisted bilayer graphene \cite{cao2018unconventional} and TMDs \cite{shimazaki2020strongly,xu2020correlated,an2020interaction,andersen2021excitons}, interlayer excitons trapped in strain-induced \cite{montblanch2021confinement} and moiré\cite{rivera2018interlayer,seyler2019signatures,baek2020highly} potentials, hybridization of excitons \cite{alexeev2019resonantly,tang2021tuning,zhang2020twist,paradisanos2020controlling} and first hints of collective phenomena such as condensation \cite{wang2019evidence,sigl2020signatures,PhysRevB.103.L041406}.\\
\indent Polarization-resolved SHG measurements (PSHG) allows determining the crystallographic orientation of monolayers, as well as the twist angle and charge transfer dynamics in bilayer heterostructures \cite{hsu2014second,jiang2014valley,psilodimitrakopoulos2019twist,lin2021twist,kim2020second,psilodimitrakopoulos2020real,zimmermann2020directional}. The SHG intensity reaches a maximum value when the electric field vector of the excitation laser beam is parallel to the armchair orientation of TMD monolayers according to the non-linear susceptibility tensor $ \chi^{2} $ of the $ D_{3h} $ space group \cite{wen2019nonlinear,klimmer2021all}. \\
\begin{figure*}
\includegraphics[width=0.9\linewidth]{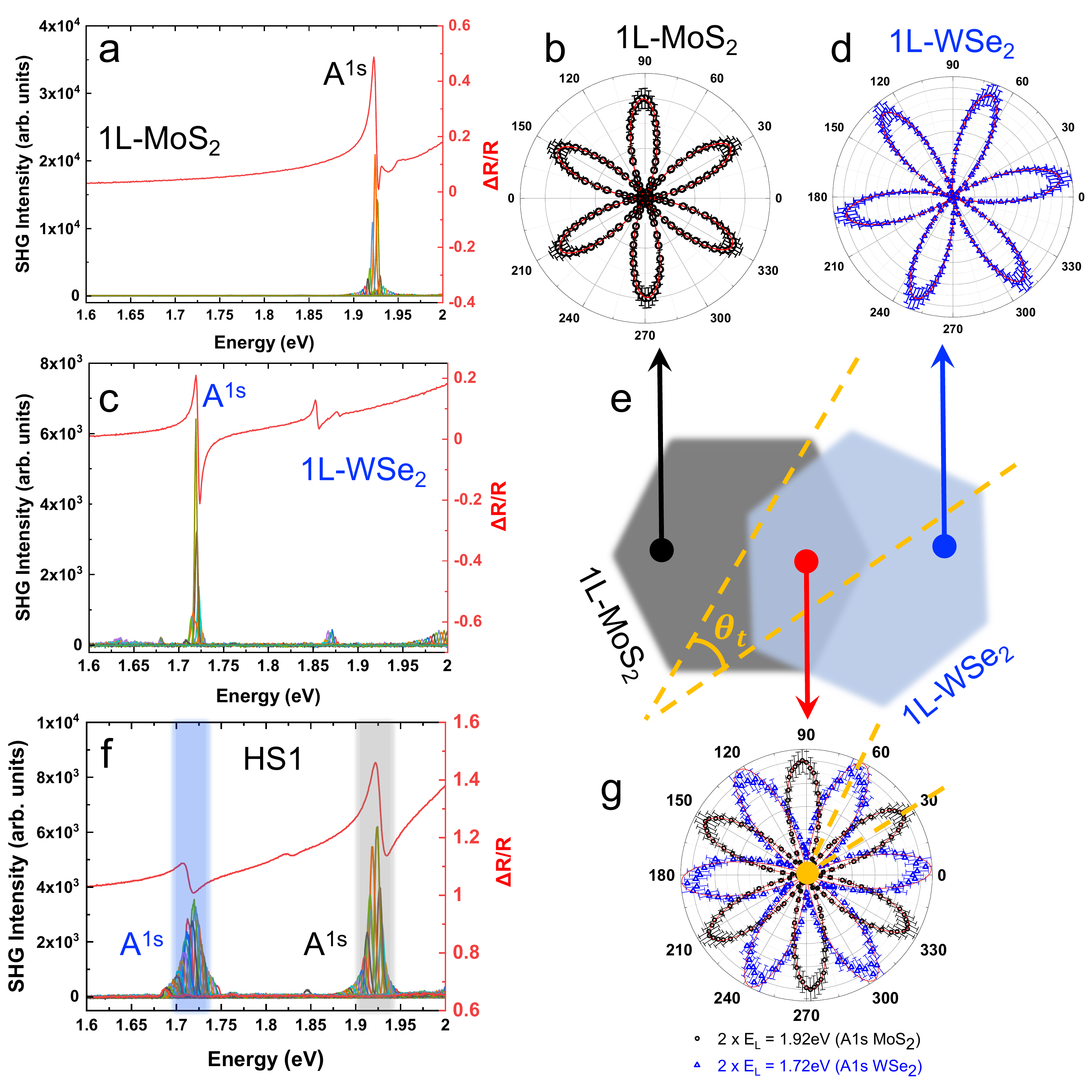}
\caption{\label{fig:fig1} \textbf{SHG spectroscopy on monolayers and heterostructure region}(a) Differential reflectivity (red) and wavelength dependent SHG of 1L-MoS$_2$.  (b)  Polar plot of SHG in 1L-MoS$_2$ at $2\times E_L=1.92~$eV. We measure over an angle range of 120 $^o$ and generate the rest of the plot by repeating the data. (c) same as (a) but for 1L-WSe$_2$. 
(d) same as (b) but at $2\times E_L=1.72~eV$ 1L-WSe$_2$. (e) schematic drawing of the different areas of the heterostructure with access also to monolayer regions. The twist angle $\theta_t$ is indicated. (f) same as (a) and (c) but for the heterostructure region. (g) Polar plots for the heterostructure region but at two \textit{different} values of $E_L$: blue triangles for $2\times E_L=1.72~$eV and black circles for $2\times E_L=1.92~$eV. All the polar plots are fitted with a cos$^{2}3(\theta-\theta_{0})$ function (red lines), where $\theta$ is the angle of the pump laser polarization and $\theta_{0}$ is a free parameter to extract the armchair orientation with respect to the lab axis (here the x-axis). }
\end{figure*}
\indent The twist angle between two monolayers in a heterostructure is usually measured by PSHG experiments on the constituent monolayers and not on the heterostructure region itself \cite{schaibley2016directional,seyler2019signatures,psilodimitrakopoulos2020real, jin2019observation, montblanch2021confinement}. This approach has several drawbacks, as monolayer drift and rotation are possible during the assembly process. In addition, strain and lattice reconstruction can naturally occur due to the interaction between the layers \cite{weston2020atomic,andersen2021excitons}, depending on the twist angle and lattice mismatch between the monolayers \cite{tran2020moire,PhysRevB.98.224102}. Direct information from the assembled bilayer region is therefore needed.
\\
\indent Here, we perform SHG spectroscopy and PSHG with a tunable laser on several MoS$_2$/WSe$_2$ heterostructures with different twist angles. Our target is to obtain information from the individual layers in the already assembled device in order to control the overall SHG response. The investigated heterobilayer system is relevant for optoelectronic applications extending to the infrared and also as a versatile platform for moiré physics \cite{pan2018quantum,kunstmann2018momentum,karni2019infrared,zhang2017interlayer,karni2021moir}. 
We measure the SHG signal amplitude as a function of laser energy $E_L$, in the energy range of A and B excitons. We tune the interference between the SHG signals from the constituent monolayers. We show that the contribution of each monolayer to the global SHG signal from the bilayer strongly depends on energy : 
For the individual  monolayers, we observe orders of magnitude enhancement of the SHG intensity when twice the laser excitation energy is tuned to the A1s-exciton of each monolayer alone (these resonances are well separated in energy : 1.72~eV for ML WSe$_2$, 1.92~eV for MoS$_2$).
When investigating the heterobilayer with a twist angle $\theta_t$, by varying the laser energy, we distinguish three main scenarios : (i) The SHG signal from the bilayer is dominated by WSe$_2$  at $2 \times E_L=1.72$~eV, as the MoS$_2$   SHG contribution at this energy is orders of magnitude weaker. (ii) At $2 \times E_L= 1.92$~eV the SHG from the heterobilayer is dominated by MoS$_2$ and the WSe$_2$ contribution is negligible. (iii) When $2 \times E_L$ is in resonance with a continuum of states from the heterobilayer bandstructure, the SHG signal of the bilayer can be seen as the interference of 2 plane waves.  This approach based on tuning $2 \times E_L$ to different exciton resonances allows accessing information on the crystallographic orientation of the individual monolayers in a heterostructure, i.e. our experiments allow to extract information on twist planes in an assembled vdW stack \cite{yao2021enhanced}.\\
\indent For the heterostructure sample with close to 0$^{o}$ twist angle we observe enhancement in the SHG intensity at very different exciton resonance energies as compared to the large twist angle sample, possibly linked to strain, the hybridization of the electronic states or  the formation of moiré minibands \cite{pan2018quantum,jin2019observation,waters2020flat,li2021imaging,cho2021highly}. We believe our results can be directly applicable to many multilayer systems with distinct exciton states \cite{tran2020moire}. \\
\begin{figure*}
\includegraphics[width=1\linewidth]{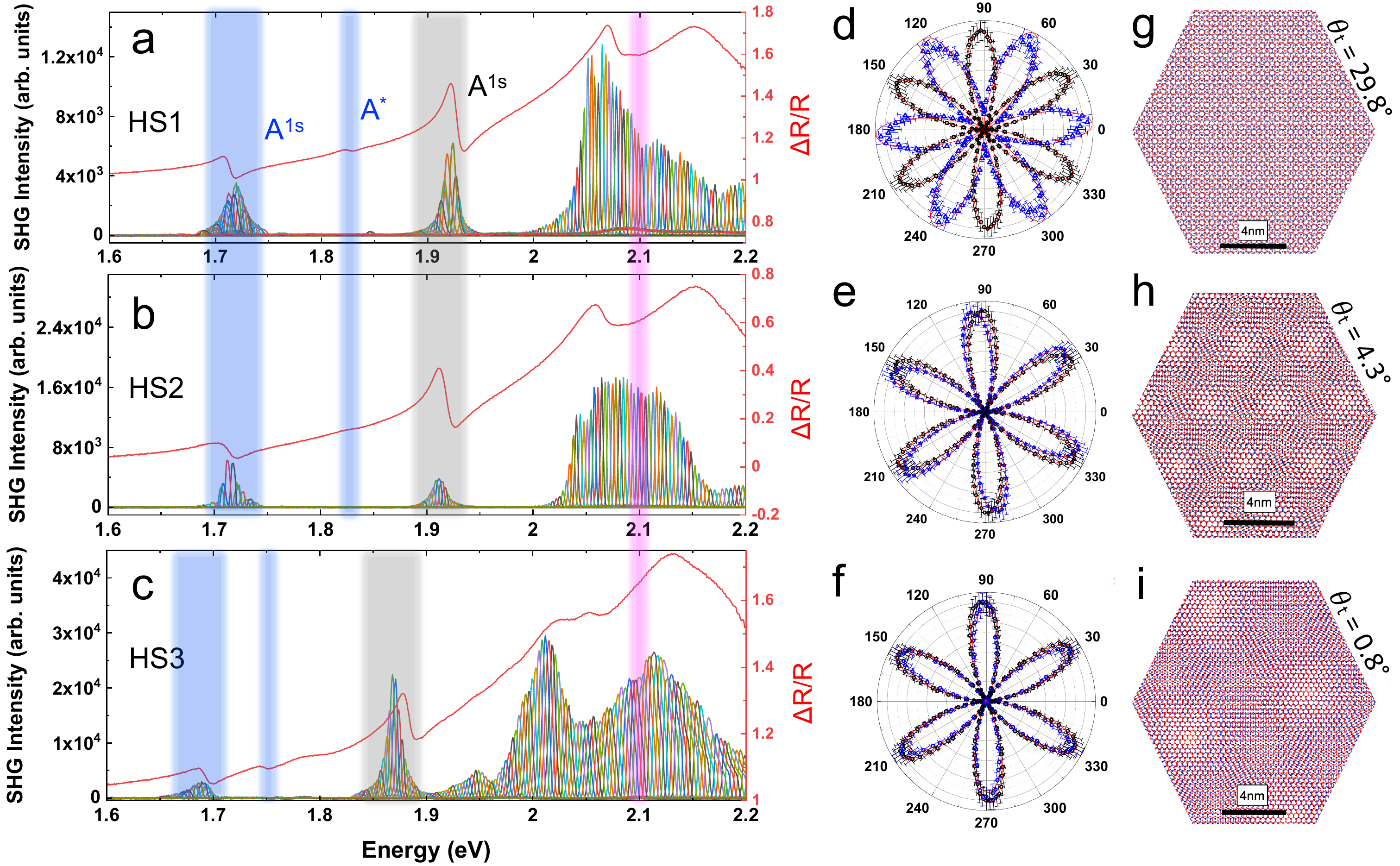}
\caption{\label{fig:fig2} \textbf{SHG spectroscopy and polarization resolved SHG for three different MoS$_2$/WSe$_2$ heterostructures.} From top to bottom: differential reflectivity (red) and wavelength dependent SHG of (a) HS1 studied in Fig.~\ref{fig:fig1}, (b) HS2, (c) HS3. The A-exciton states of MoS$_2$ and WSe$_2$ are indicated with black and blue shaded regions, respectively. The respective polar plots when the twice the excitation energy is in resonance with the corresponding A-excitons of MoS$_2$ (black) and WSe$_2$ (blue) are shown in (d),(e),(f). Schematic representation of the moiré pattern of 1L-MoS$_2$/1L-WSe$_2$ heterostructures with twist angles $\theta_t$ of (g) HS1 = 29.8$ ^{o} $, (h) HS2 = 4.3$ ^{o} $  and (i) HS3 = 0.8$ ^{o} $. The extracted moiré period is 0.62~nm, 3.84~nm and 8.45~nm, respectively. Regions with main excitonic transitions are shaded blue and gray. A magenta stripe is shown at 2.1 eV to indicate $2 \times E_L$ used for the experiments presented in Fig.~\ref{fig:fig3} }
\end{figure*}

\textbf{RESULTS}\\
We investigate three heterobilayers HS1-3 with different twist angles $\theta_t$ between WSe$_2$ and MoS$_2$. The structures share the same MoS$_2$ monolayer and the same hBN encapsulation (top and bottom) to allow isolating changes due to the twist angle only, see schematics in Fig.~\ref{fig:fig1}e and supplement A. For SHG spectroscopy we use a pulsed Ti-Sapphire laser source coupled to an optical parametric oscillator (OPO) and we scan twice the laser energy ($i.e.$ 2$\times$E$_{L}$) over the energy range that corresponds to the main optical transitions. Experiments are performed at a temperature of T = 5 K in vacuum in a confocal microscope (excitation/detection spot diameter of the order of the wavelength), see \cite{shree2020guide} and Methods . \\
\indent \textbf{SHG comparison monolayers versus heterobilayers.---}
\indent In Fig.~\ref{fig:fig1}a,c we plot SHG spectra from the MoS$_2$ and WSe$_2$ monolayers for laser energies between 2$\times$E$_{L}$ = 1.6 to 2~eV in steps of $\approx$ 3~meV between two adjacent spectra. Each SHG spectrum is composed of a single peak that corresponds to a separate data acquisition for each excitation laser energy E$_{L}$ (see supplement B) \cite{shree2021interlayer}. For 1L-MoS$_2$ the SHG amplitude is strongly enhanced at 1.92~eV, for 1L-WSe$_2$ at 1.72~eV.
Comparing SHG spectra with white light reflectivity plotted on the same panels in  Fig.~\ref{fig:fig1}a,c, we conclude that the maxima in the SHG amplitude occur when 2$\times$E$_{L}$ is resonant with intralayer A excitons ($i.e.$ Coulomb-bound electron-holes pairs within the same layer) \cite{gerber2019interlayer,wang2015giant}. The excitonic contribution to the SHG intensity can be orders of magnitude higher than the intrinsic contribution from the crystal, as reported  for TMD monolayers and other material systems \cite{seyler2015electrical,wang2015exciton,glazov2017intrinsic,zhao2016atomically,Yakovlev:2018aa,abdelwahab2018highly,lafeta2021second}. \\
\indent For the MoS$_2$ monolayer we perform PSHG at the A exciton resonance (2$\times$E$_{L}$ = 1.92 eV) to extract the crystallographic orientation, see Fig.~\ref{fig:fig1}b. We measure the SHG amplitude as a function of the linearly polarized excitation angle with respect to the in-plane crystallographic orientation (see supplement C) and we present the results in a polar plot in Fig.~\ref{fig:fig1}b.
We repeat the corresponding experiment for the WSe$_2$ monolayer by tuning now the laser energy in resonance with the corresponding A-exciton (2$\times$E$_{L}$ = 1.72 eV) and perform PSHG, see polar plot in Fig.~\ref{fig:fig1}d. For both PSHG measurements on monolayers the measured SHG integrated intensity modulation can be fitted with a  cos$^{2}3(\theta-\theta_{0})$ function, where $\theta$ is the angle of the laser polarization and $\theta_{0}$ is a free parameter to extract the armchair orientation with respect to the lab axis (here the x-axis) \cite{li2013probing}. Based on the relative armchair orientation, we determine a twist angle  ($\theta_t =$21.45 $ \pm $ 0.39)$^{o}$ between the individual monolayers (supplement C for discussion on the error bars).\\
\indent For the measurements in Fig.~\ref{fig:fig1}f, we place the excitation/detection spot on the heterobilayer region of HS1 where MoS$_2$ and WSe$_2$ overlap, see red arrow in Fig.~\ref{fig:fig1}e., and we perform SHG spectroscopy.
The results in Fig.~\ref{fig:fig1}f, show strong resonances at the energies of the intralayer A-exciton of WSe$_2$ (1.72~eV) and MoS$_2$ (1.92~eV), respectively. Note that the overall nonlinear response of the A-excitonic resonances appears broader in the heterostructures (full width at half maximum, FWHM $ \approx 15 $~meV, see Fig.~\ref{fig:fig1}f) compared to the bare monolayers (FWHM $ \approx 5 $~meV, see Fig.~\ref{fig:fig1}a,b). This can be due to fast charge transfer processes due to the type-II band alignment \cite{hong2014ultrafast} or/and disorder (i.e. strain, impurities) introduced by the additional transfer steps.\\
\indent In a second step, we tune $2 \times E_L$ on the WSe$_2$ intralayer exciton resonance and we perform PSHG, see blue data points in Fig.~\ref{fig:fig1}g. Then we change the laser energy to $2 \times E_L = 1.92~$eV and perform  PSHG experiments, see black data points in Fig.~\ref{fig:fig1}g. Comparing the two data sets we see two very distinctive polar plots using two different values of $E_L$, although we carried out the measurements at the same spot of the heterobilayer. We extract a twist angle of $\theta_t =$(29.82 $ \pm $ 0.41)$^{o}$ between the two polar plots in Fig.~\ref{fig:fig1}g, an angle slightly larger than determined for the separate monolayer orientations, possibly indicating a rotation of the WSe$_2$ layer during the assembly process, see supplement D.
If we compare the PSHG from the heterobilayer with the monolayer measurements, we find that for the heterostructure, the WSe$_2$ contribution totally dominates at $2 \times E_L =1.72~$eV, whereas at $2 \times E_L =1.92~$eV the MoS$_2$ contribution is predominant, see more quantitative analysis in the discussion section. The intralayer A-exciton states of WSe$_2$ and MoS$_2$ have a distinct contribution in the SHG response of the heterostructure. As a result, for certain energies one of the two constituent monolayers will exclusively contribute to the SHG signal. So by varying the laser energy on an already assembled heterostructure we can determine $in-situ$ the crystallographic direction of the constituent layers.  This has important implications, as the twist angle determines bandstructures and hence electrical and optical properties. \\
\indent We did not scan the lower energy range for interlayer excitons, as in SHG spectroscopy and linear absorption mainly direct transitions with high oscillator strength are visible.\\
\indent \textbf{SHG spectroscopy for heterobilayers with different twist angle.---}
\indent Our next target is to investigate the SHG response as a function of twist angle as shown in Fig.~\ref{fig:fig2} for three different heterostructures. We show in Fig.~\ref{fig:fig2}a an extended scan of the results from Fig.~\ref{fig:fig1}f on HS1. We reveal at energies $2 \times E_L > 2.05~$eV an onset of a continuum of states. In Fig.~\ref{fig:fig2}b we cover the same energy range but investigate a different WSe$_2$ layer on top of MoS$_2$ in HS2, see supplement A for sample images. We perform for HS2 PSHG at $2 \times E_L = 1.72~$eV  (blue data in Fig.~\ref{fig:fig2}e) and $2 \times E_L = 1.92~$eV  (black data in Fig.~\ref{fig:fig2}e). Also for HS2 we obtain two very different angle dependencies at these different laser energies. This allows to extract \textit{in-situ} a twist angle between the layers of $\theta_t =$(4.33 $ \pm $ 0.51)$^{o}$, in very good agreement with the twist angle extracted from the orientation of the monolayer parts outside the heterostructure region, see supplement D. The SHG spectroscopy results obtained throughout these studies for HS1 and HS2 are very consistent in terms of exciton resonances and continuum onset, the different twist angle $\theta_t$ is the most striking difference between HS1 and HS2. \\
\indent In Fig.~\ref{fig:fig2}g,h and i we plot monolayer WSe$_2$ on top of monolayer MoS$_2$ for different twist angles in an idealized lattice configuration (ignoring lattice reconstruction). By considering the lattice mismatch between MoS$_2$ and WSe$_2$, we graphically extract the periodicity of the moiré superlattice, to be 0.62~nm and 3.8~nm  for twist angles that correspond to HS1 and HS2, respectively. Interestingly, due to the large lattice mismatch of 3.7\% also zero degree twist angle results in principle in the formation of a periodic moiré landscape (period $\approx 8.5~$nm. Quantum-confined electronic states and moiré patterns in MoS$_2$/WSe$_2$ heterobilayers have been observed previously using spectroscopy techniques and electron microscopy \cite{zhang2017interlayer,pan2018quantum}. \\
\indent For HS3 with close to zero twist angle, we measure a very different response in linear and non-linear optics as compared to the larger twist angle samples, namely : (i) in both differential reflectivity and SHG spectroscopy the main resonances are strongly red-shifted by 40~meV (ii) for HS3 in the energy range $2 \times E_L$ from 1.9 to 2.05~eV strong SHG signal amplitude is recorded. In contrast, over this energy range the signal for the other structures HS1 and HS2 is close to zero. Both observations indicate that the band-structure for HS3 is altered as compared to HS1 and HS2. In order to perform PSHG on intralayer excitons, we tune the laser energy to the red-shifted resonance energies at 1.68~eV and 1.88~eV clearly visible in Fig.~\ref{fig:fig2}c, the corresponding angle dependent data is plotted in  Fig.~\ref{fig:fig2}f. As expected for a sample with close to zero (or 60$^{o}$) twist angle, the two sets of data are oriented along the same axis. As for HS3 we find an SHG signal amplitude that is in general roughly equal to the sum of the measured monolayer SHG amplitudes (for $2 \times E_L = 2.1~$eV), we conclude that the zero degree twist angle corresponds to R-type and not H-type (i.e. 60$^{o}$) stacking \cite{hsu2014second}. 
\\
\begin{figure}
\includegraphics[width=0.8\linewidth]{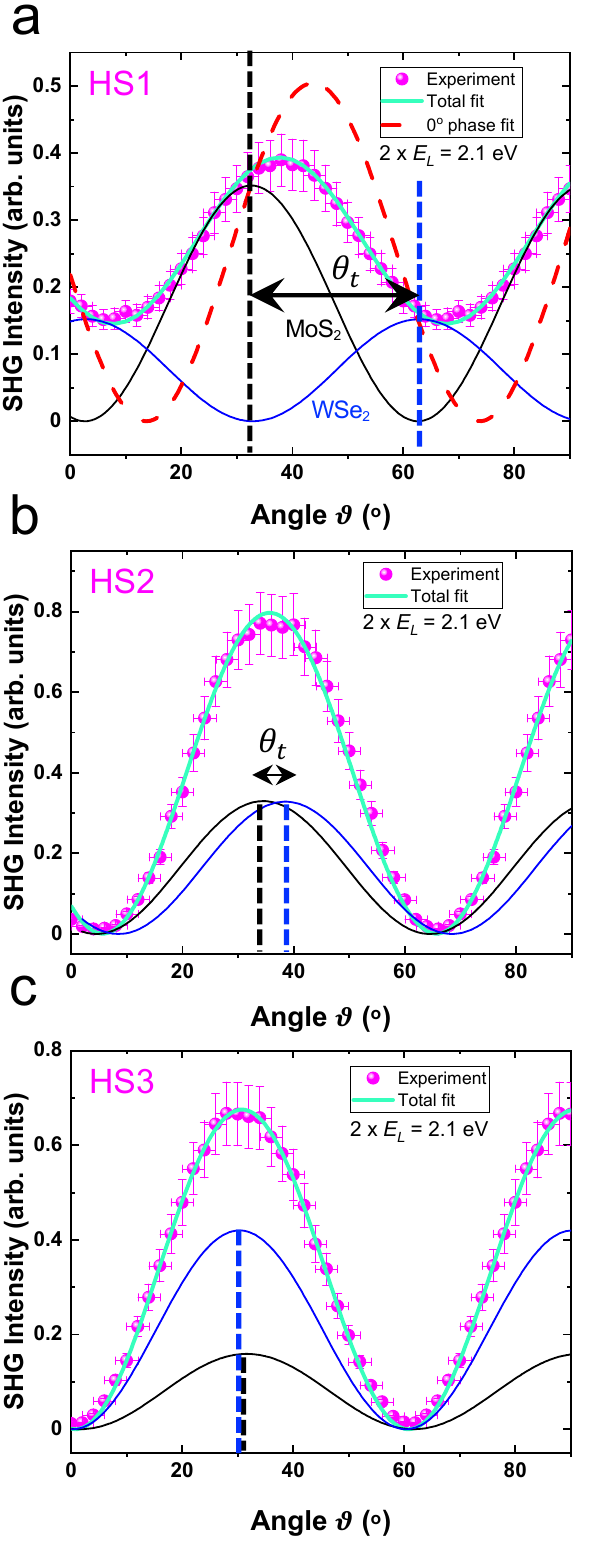}
\caption{\label{fig:fig3} \textbf{Polarization angle dependent SHG and fitting}. Using higher excitation energies (2$\times$E$_{L}$= 2.1 eV for (a) HS1, (b), HS2 and (c) HS3. Experimental points are presented with magenta spheres and the fitting of the total SHG signal is shown in green lines. The twist angle values $\theta_t$ used in Eq.\ref{Ipphase} for the total fit are the experimental values $ 29.8^{\circ},  4.3^{\circ}$ and $ 0.8^{\circ} $ for HS1, HS2 and HS3, respectively. The amplitude ratio $L_1$(MoS$_2$) : $L_2$(WSe$_2$) is 1.5, 1 and 0.6, while the phase, $ \varphi $, is kept $79^{\circ}$ in all cases. The SHG modulation of the individual layers that construct the total SHG is shown in black and blue lines for MoS$_2$ and WSe$_2$, respectively. For HS1 (a) where there is a finite minimum to maximum ratio due to the phase shift and large twist angle, a fitting with zero phase shift is shown in dashed red line for comparison. }
\end{figure}
\indent \textbf{SHG polarization control.---}
\indent In addition to the main intralayer exciton resonances, we also investigated the SHG response at other energies that contain crucial information on the heterostructure. In Fig.~\ref{fig:fig3} we plot the PSHG results for the measurements at $2 \times E_L=2.1~$eV in the continuum of states for HS1-3, indicated by a magenta stripe in Fig.~\ref{fig:fig2}a-c. The results are very different compared to the SHG close to intralayer exciton resonances : for HS1 we find that the signal is maximized along an axis that is neither aligned with MoS$_2$ nor with WSe$_2$, compare black and blue lines with magenta spheres in Fig.~\ref{fig:fig3}a. Here, the black and blue dashed lines correspond to the armchair orientation of MoS$_2$ and WSe$_2$, respectively. A striking feature is that the signal does not go to zero, so we have a strongly changed minimum to maximum SHG signal ratio as a function of the polarization angle for this measurement. This is signature of elliptically polarized SHG emission \cite{hecht2017optics,kim2020second}, in contrast to the linearly polarized SHG plotted in Fig.~\ref{fig:fig2}d-f of the same heterobilayer sample but at lower laser energy. \\
\indent For HS2  PSHG measurements also result in a slightly shifted polar plot as compared to the monolayer orientations but now the minimum to maximum SHG signal ratio is close to zero indicating close to linear polarization. From our measurements we determine an angle for the polarization maximum in-between the MoS$_2$ and WSe$_2$ armchair directions. \\
\indent For HS3 close to zero twist angle ($\theta_t =$0.79$^{o}$ $ \pm $ 0.53$^{o}$) the PSHG results show linearly polarized light, and not elliptically polarized light. For the three heterostructures at $2 \times E_L=2.1~$eV, we use the twist angle, individual layer contributions and phase differences to fit the measured data with a very simple plane wave model \cite{kim2020second}, see discussion section.  \\

\textbf{DISCUSSION} \\
\indent First, we discuss for HS3 with $\theta_t$ = 0.8$^{o}$ twist angle possible origins of the redshift of the main intralayer excitonic transitions and the appearance of new resonances, which still need further investigation. Comparing with the literature, our observations are in agreement with emergent moiré superlattice exciton states when the two lattices are closely aligned. This has been previously observed using reflectivity and photoluminescence excitation (PLE) experiments in WS$_2$/WSe$_2$ heterostructures \cite{jin2019observation} where the lattice mismatch of 4$\%$ is very similar to MoS$_2$/WSe$_2$ with 3.7$\%$.  In a study that combines spectroscopy with density functional theory (DFT), a large in-plane strain variation across the moiré unit cell of MoS$_2$/WSe$_2$ heterobilayer has been proposed \cite{waters2020flat}, leading to energy shifts. The important role of strain for MoS$_2$/WSe$_2$ heterobilayer is also shown when investigating interlayer exciton formation \cite{cho2021highly}. In reconstructed WSe$_2$/WS$_2$ moiré superlattices quantitative studies of moiré flat bands are reported by comparing scanning tunnelling
spectroscopy of high-quality exfoliated TMD heterostructure devices with ab initio simulations of TMD moiré superlattices. Also in these studies large in-plane strain redistribution is identified in WSe$_2$/WS$_2$ moiré heterostructures linked to 3D buckling \cite{li2021imaging}. \\
\indent Second, we analyze the main experimental findings for the samples with non-zero twist angle, HS1 and HS2. The main observations are linearly polarized SHG for the specific case of $2 \times E_L$ in resonance with the intralayer exciton resonances, where the polarization axis is aligned with the airmchair direction of the monolayer whose exciton resonance was addressed. We then show that at higher laser energies the SHG from the heterostructure is elliptically polarized for HS1 and close to linearly polarized for HS2 , with a polarization axis away from the monolayer orientations.\\
\indent An interesting analysis of SHG from \textit{homobilayers} with different twist angle is given in Ref. \citep{hsu2014second}. The situation is more complex in \textit{heterobilayers}, where both monolayers have finite, yet different contributions in the total SHG signal,
as discussed in \cite{kim2020second}. As a result, the minimum to maximum SHG signal ratio (i.e. minor to major axial ratio of polarization, $\rho$) can change considerably for large twist angles (e.g. see Fig.\ref{fig:fig3}a). The combination of a phase difference between the two fields and a finite twist angle introduce an elliptical polarization in the total field \cite{hecht2017optics}. To go further in our analysis, we adopt the simple approach proposed in \cite{kim2020second} based on the interference of plane waves, which does not explicitly include any layer hybridization. In a PSHG experiment on a heterobilayer, the parallel component of the SHG signal can be expressed as : 
\begin{equation}
\label{Ipphase}
\begin{split}
I_{P}  & = L_1^{2}\cos^{2}(3\vartheta)+L_2^{2}\cos^{2}[3(\theta_{t} - \vartheta)] \\
& + 2 L_1 L_2\cos(3\vartheta) cos[3(\theta_{t} - \vartheta)]\cos\varphi 
\end{split}
\end{equation}
where $L_{1,2}$ are the SHG amplitudes of the first and second monolayer, $ \vartheta $ is  the relative angle between the fundamental (Laser) polarization and the armchair direction of the monolayer, $ \theta_{t} $ is the twist angle and $ \varphi $ the phase difference between the two SHG fields. In our experiment we vary $ \vartheta $ by rotating a superachromatic half-wave plate, see supplement C. \\ 
\indent In the experiments presented here, we bring together very specific conditions that allow us to vary SHG signals from a heterobilayer described by Eq.\ref{Ipphase} : (i) diffraction limited detection/excitation spot size, so we can compare monolayer and heterostructure response without spatial overlap as the spot size is smaller than the lateral dimensions of the sample regions (ii) our tunable source is a ps-laser and has a narrow spectral FWHM of typically 4~meV, (iii) we perform experiments at low temperature in high quality hBN encapsulated structures, which results in spectrally narrow (few meV FWHM) exciton transition linewidth \cite{cadiz2017excitonic} (see details in supplement A) and allows addressing individual optical resonances in the two layers. (iv) A-exciton resonances in WSe$_2$ and MoS$_2$ lie spectrally 200~meV apart, allowing to target them separately. \\
\indent By carefully adjusting experimental parameters (i) to (iv) we can control the SHG signal of the heterostructure. \textit{Linear polarization} of the SHG signal from the heterobilayer can be occur for several configurations, namely (1)  by tuning the $2 \times E_L$ into resonance with the A-exciton state of one of the two monolayers. Independently of the twist angle then $ I_{P}$ reduces to a single cos$ ^{2}(3 \theta)$ term since the SHG from the selected monolayer will dominate and either $L_1$ or $L_2$ are zero in Eq.~\ref{Ipphase}. This allows to describe the results in Fig.~\ref{fig:fig2}e and f and also Fig.~\ref{fig:fig1}g. (2) at energies where both monolayers contribute to the SHG signal (i.e. both $L_1$ and $L_2\neq 0$) and for non-zero twist angles, linear polarization occurs if the phase difference $ \varphi =0$, see red dashed curve in Fig.~\ref{fig:fig3}a. (3) in the case of aligned heterostructures (see HS3 with $\theta_t\simeq$ 0$^{o}$) the polarization of the total SHG field is always linear.\\
\indent In order to generate \textit{elliptical polarization} as observed for HS1 both the twist angle $ \theta_{t}$ and the phase shift $ \varphi$ need to be non-zero in Eq.~\ref{Ipphase}. In the polarization dependent plots as a function of $ \vartheta $ in Fig.~\ref{fig:fig3}a  elliptical polarization results in a different $\rho$. For instance, in HS1 the twist angle is large ($ \approx $ 30$ ^{o} $) and the minimum to major axial ratio is one order of magnitude different when we compare 2$ \times E_{L} $ = 1.92 eV (Fig.\ref{fig:fig2}d, $\rho\simeq$ 0.03:1) with respect to 2$ \times E_{L} $ = 2.1 eV (Fig.\ref{fig:fig3}a, $\rho\simeq$ 0.39:1). For the fits in Fig.~\ref{fig:fig3}a,b, and c based on Eq.~\ref{Ipphase} we use the twist angle determined beforehand for each HS. We treat the amplitudes $L_1$ and $L_2$ and the phase $\varphi$ as fitting parameters in HS1 and then we use the same phase for HS2 and HS3, the results are summarized in the caption of Fig.~\ref{fig:fig3}. Our target here is not to quantify the phase, $\varphi$, but to demonstrate that the total SHG signal from a twisted heterostructures can be either linearly or elliptically polarized, depending on whether both monolayers contribute to the overall signal. Spectral phase interferometry can be employed to measure $\varphi$, as has been demonstrated elsewhere (see Ref. \cite{kim2020second}). The \textit{elliptical polarization} we experimentally obtain for HS1 at 2$\times E_L  = 2.1$~eV indicates exciton resonances from either layer at this energy which have a different phase \cite{chang1965relative}. This could possibly be explained by different contributions from intralayer B-excitons of WSe$_2$ and MoS$_2$ monolayers at this energy \citep{kormanyos2015k,robert2018optical,stier2018magnetooptics}. \\
\indent For small and close to zero twist angles of HS2 and HS1 the experimentally measured polarization of the SH is close to linear as expected from Eq.~\ref{Ipphase}, so fit parameters are not uniquely defined, as varying the amplitude ratio and the phase have numerically the same effect on the overall polarization and amplitude.\\

\textbf{CONCLUSIONS} \\
We have performed PSHG spectroscopy in MoS$_{2}$/WSe$_{2}$ heterostructures. Tuning twice the excitation energy in resonance with an intralayer excitonic transition of the constituent monolayers allows the $in-situ$ determination (directly on the heterostructure) of the twist angle. This is a consequence of the SHG amplitude and polarization control by addressing energetically distinct intralayer exciton resonances in the top and bottom layer. This approach can be applied to a large number of heterobilayer systems \cite{tran2020moire}. We conclude that the total SHG intensity and polarization for a given TMD heterostructure depend on (i) the twist angle, (ii) the relative amplitude between the SH fields of the constituent monolayers at a given excitation energy, and (iii) the phase difference between the SH waves. Finally, we show in SHG spectroscopy that the bandstructure of aligned (close to 0$^{o} $) MoS$_{2}$/WSe$_{2}$ heterostructures is strongly altered as compared to samples with larger twist angles. We observe considerable shifts of exciton resonance energies and the appearance of new resonances in the linear and non-linear susceptibilities. \\

\textbf{Methods.}\\
Sample : an optical microscope image of the sample is shown in the supplement A. An exfoliated monolayer MoS$_2$ with large lateral size ($\approx$ 120~$\mu$m, blue-dashed lines) lies on top of a $\approx$150~nm thick hBN, while three different exfoliated WSe$_2$ monolayers (red-dashed lines) are deliberately transferred on top of MoS$_2$ in different twist angles after alignment of the long edges of the flakes. As a result, three different heterostructures are formed, namely HS1, HS2 and HS3. In all cases, there is optical access to the bare monolayers to confirm the validity of the results collected from the heterostructures. The same, thin ($\approx$ 10~nm) top-hBN covers the whole structure. The uniformity of the top and bottom hBN thickness is important because thin-film interference effects can modify the SHG intensity and the reflectivity shape/amplitude when comparing different samples \cite{robert2018optical}. 

\textbf{Acknowledgements.}  \\
Toulouse acknowledges funding from ANR 2D-vdW-Spin, ANR MagicValley, ANR IXTASE, ANR ATOEMS, and the Institut Universitaire de France. Growth of hexagonal boron nitride crystals was supported by the Elemental Strategy Initiative conducted by the MEXT, Japan ,Grant Number JPMXP0112101001, JSPS KAKENHI Grant Number JP20H00354 and the CREST(JPMJCR15F3).\\

\newpage
\section{Supplement}

\subsection{A. Optical microscope image and SHG optical setup}

\begin{figure*}
\centerline{\includegraphics[width=140mm]{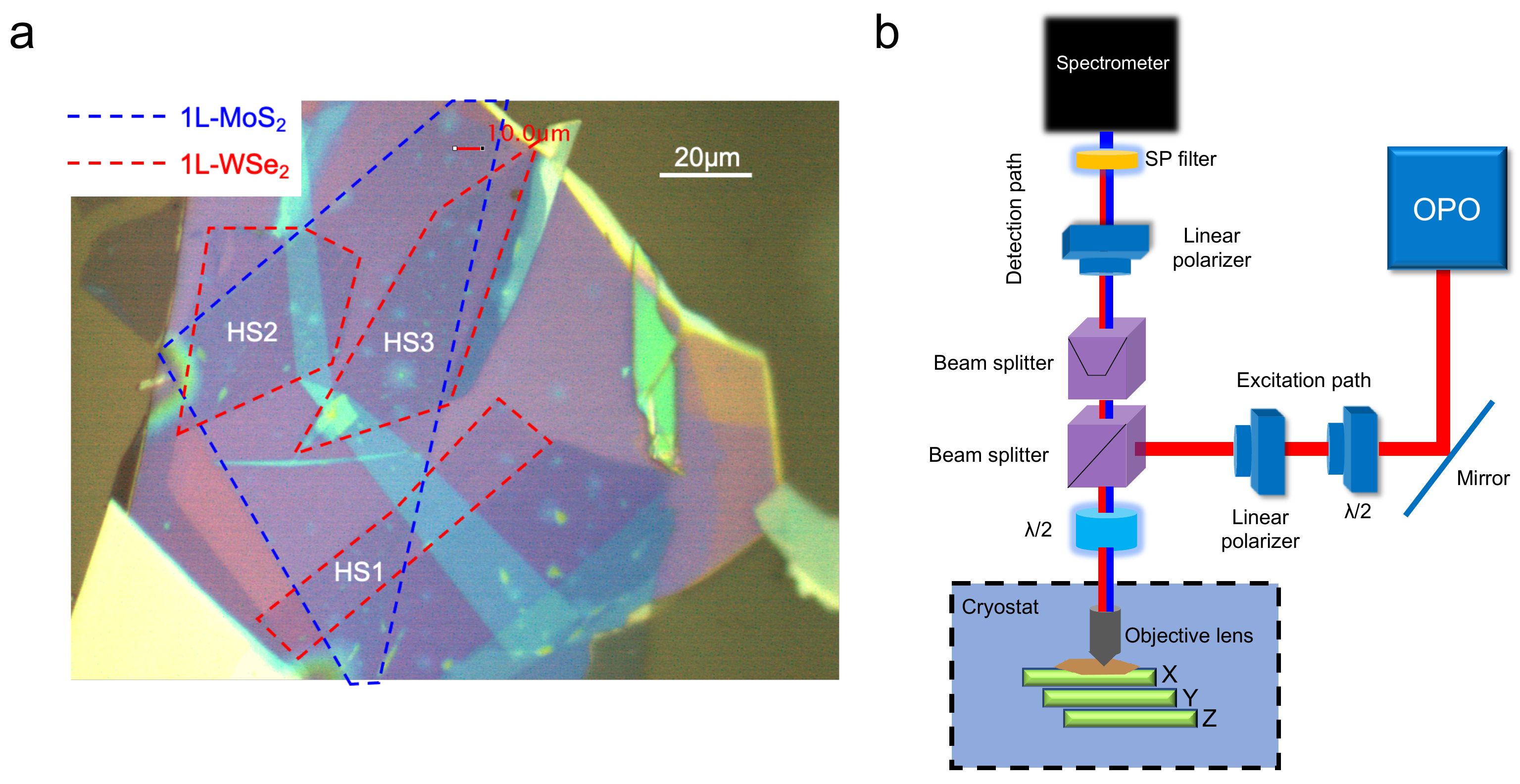}}
\caption{\label{fig:figS1} (a) Optical microscope image of the three heterostructures, HS1, HS2 and HS3. (b) Schematic of the optical setup.}
\end{figure*}

\begin{figure*}
\centerline{\includegraphics[width=140mm]{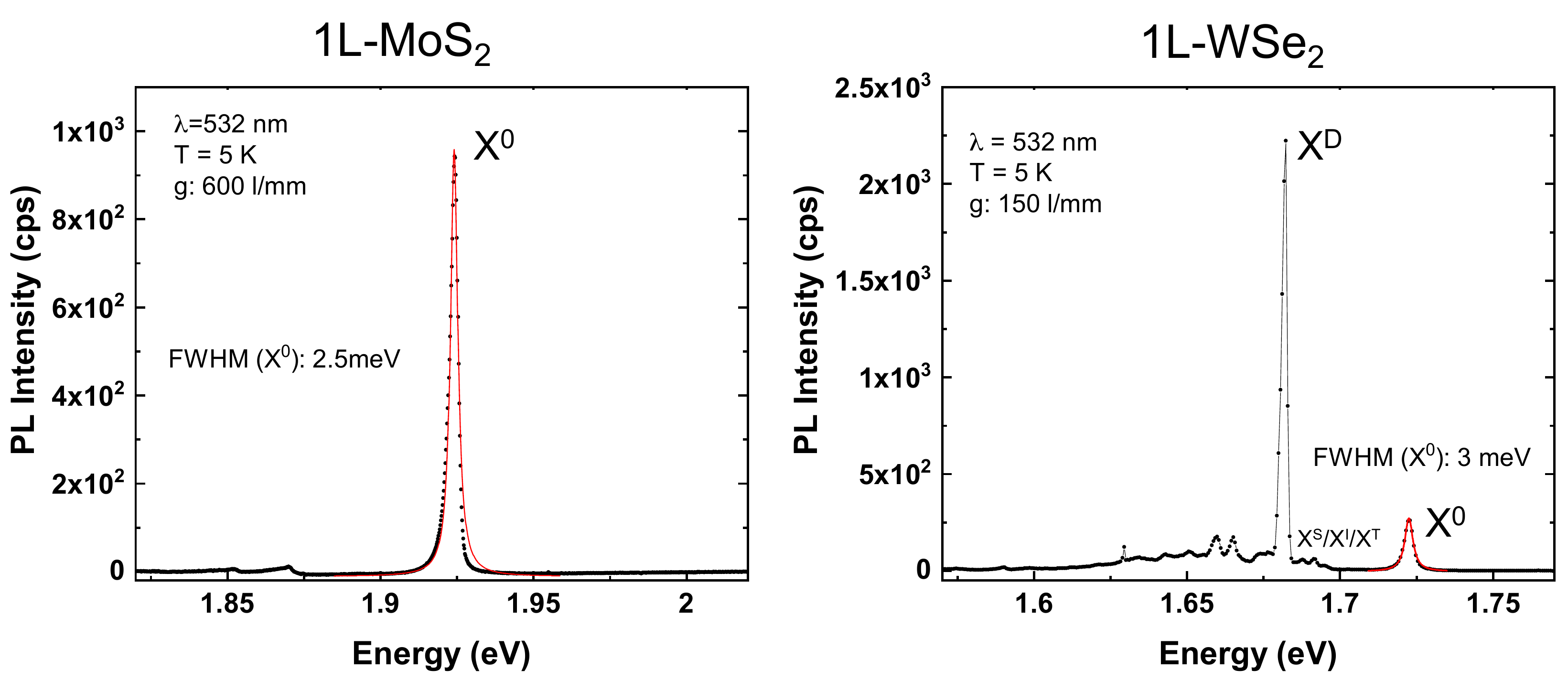}}
\caption{\label{fig:figSA} Photoluminescence spectroscopy of bare 1L-MoS$_2$ (left) and 1L-WSe$_2$ (right) collected at T = 5 K. The FWHM of the neutral exciton is 2.5 meV and 3 meV for MoS$_2$ and WSe$_2$, respectively.}
\end{figure*}

Fig.~\ref{fig:figS1}a shows an optical microscope image of the sample. Access to bare monolayers is possible in all cases, while all flakes are encapsulated by the same top and bottom hBN thickness. A large MoS$_2$ monolayer (blue, dashed lines) is first transferred on top of a $\approx$130~nm hBN. Three different WSe$_2$ monolayers (red, dashed lines) are subsequently transferred on MoS$_2$ with various twist angles. A schematic of the optical setup used for the PSHG spectroscopy experiments is shown in Fig.~\ref{fig:figS1}b. An Optical Parametric Oscillator (OPO) is aligned to a home-built confocal microscope in a closed-cycle cryostat system. A combination of linear polarizers and halfwave plates allows the control of excitation and detection polarization for the PSHG measurements. The light is focused onto the sample using a microscope objective (NA=0.75) while the temperature of the sample is T = 5 K. Cryogenic nanopositioners (nm steps, mm range) are used to control the position of the sample with respect to the laser beam. The back-reflected light from the sample is dispersed in a spectrometer with a 150 g/mm grating. The spectra are recorded by a liquid-nitrogen cooled charged coupled device (CCD) array. Low temperature reflectivity experiments were performed using a halogen lamp as a white-light source with a stabilized power supply focused initially on a pin-hole that is imaged on the sample. The excitation/detection spot diameter is $ \approx $ 1$ \mu $m, i.e. smaller than the typical diameter of the sample, see \cite{shree2020guide} for further details. Typical photoluminescence (PL) spectra of the bare MoS$_2$ and WSe$_2$ monolayers were collected after stacking, see Fig.~\ref{fig:figSA}. The PL spectra reveal good optical quality of the constituent monolayers with FWHM of 2.5 meV and 3 meV for the neutral exciton of MoS$_2$ and WSe$_2$, respectively.

\subsection{B. SHG spectroscopy}

\begin{figure}
\centerline{\includegraphics[width=80mm]{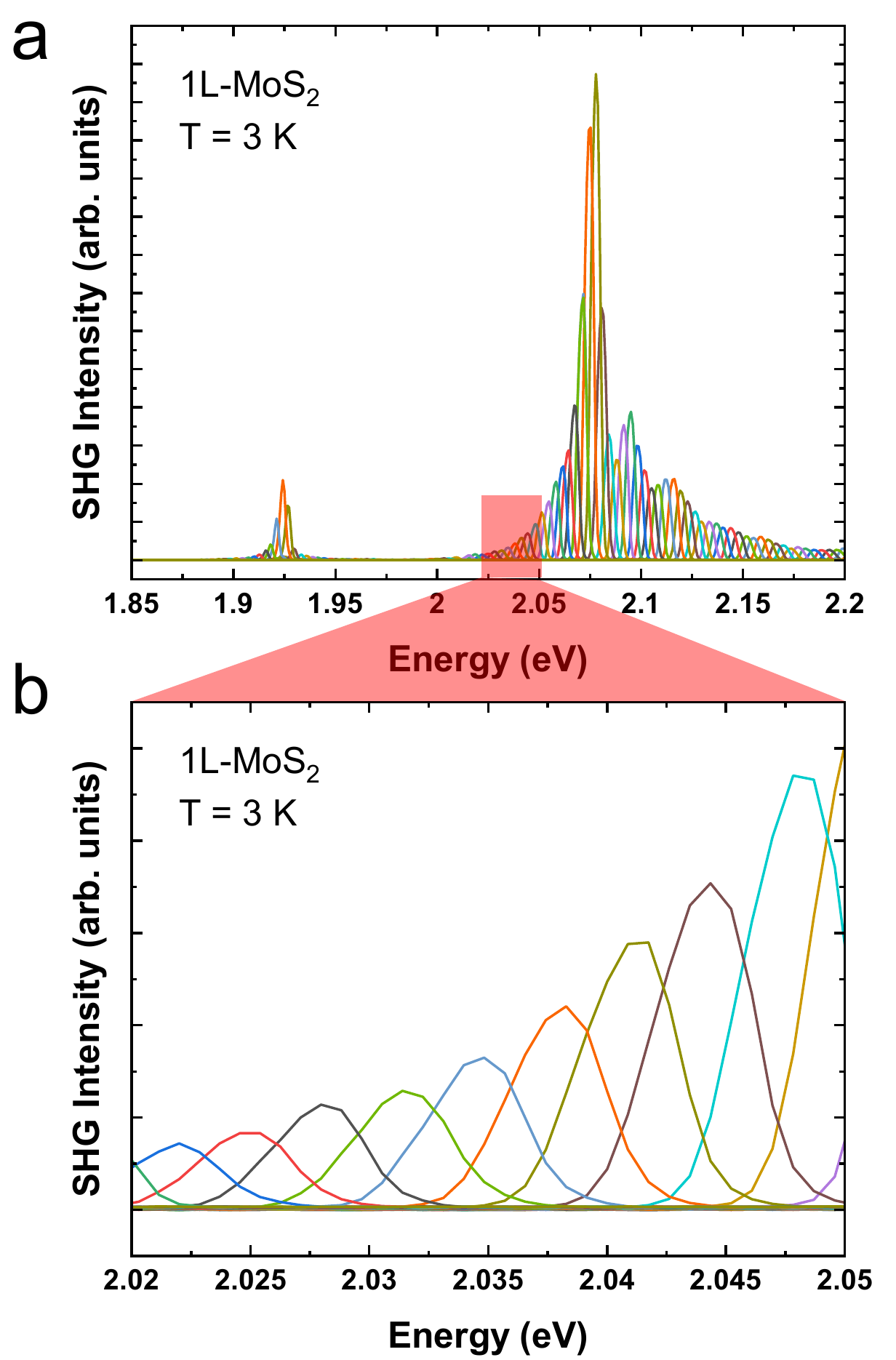}}
\caption{\label{fig:figS2} (a) Example of SHG spectroscopy of a monolayer MoS$_2$ measured at T = 3 K. A  randomly selected spectral area (red color) is zoomed in (b) to demonstrate the separate SHG experiments that construct the complete spectrum.}
\end{figure}

All the SHG spectra shown in the main text are composed of several, different SHG experiments. An example of a complete SHG spectrum of a monolayer MoS$_2$ in a range from 1.85~eV up to 2.2~eV is shown in Fig.~\ref{fig:figS2}a. Different resonances can be identified when twice the excitation energy matches the exciton energy. In a typical experiment we select the average power of the excitation laser to be 5~mW. We keep the average power stable, as well as identical acquisition parameters and we change the excitation energy in the OPO in equal steps. At each excitation step we collect a different spectrum. In Fig.~\ref{fig:figS2}b, a zoomed spectral area of the red rectangle of Fig.~\ref{fig:figS2}a, is presented. The different SHG spectra with an energy difference of $ \approx $ 3~meV can be distinguished. The pulse duration of the excitation laser is $ \approx $ 1~ps. A background subtraction needs to be applied in every step of the signal collection to exclude any external contributions in the SHG intensity. The contribution of hBN to the SHG signal is negligible \cite{li2013probing,shree2021interlayer}

\subsection{C. Polarization-resolved SHG and error bars}

\begin{figure*}
\centerline{\includegraphics[width=110mm]{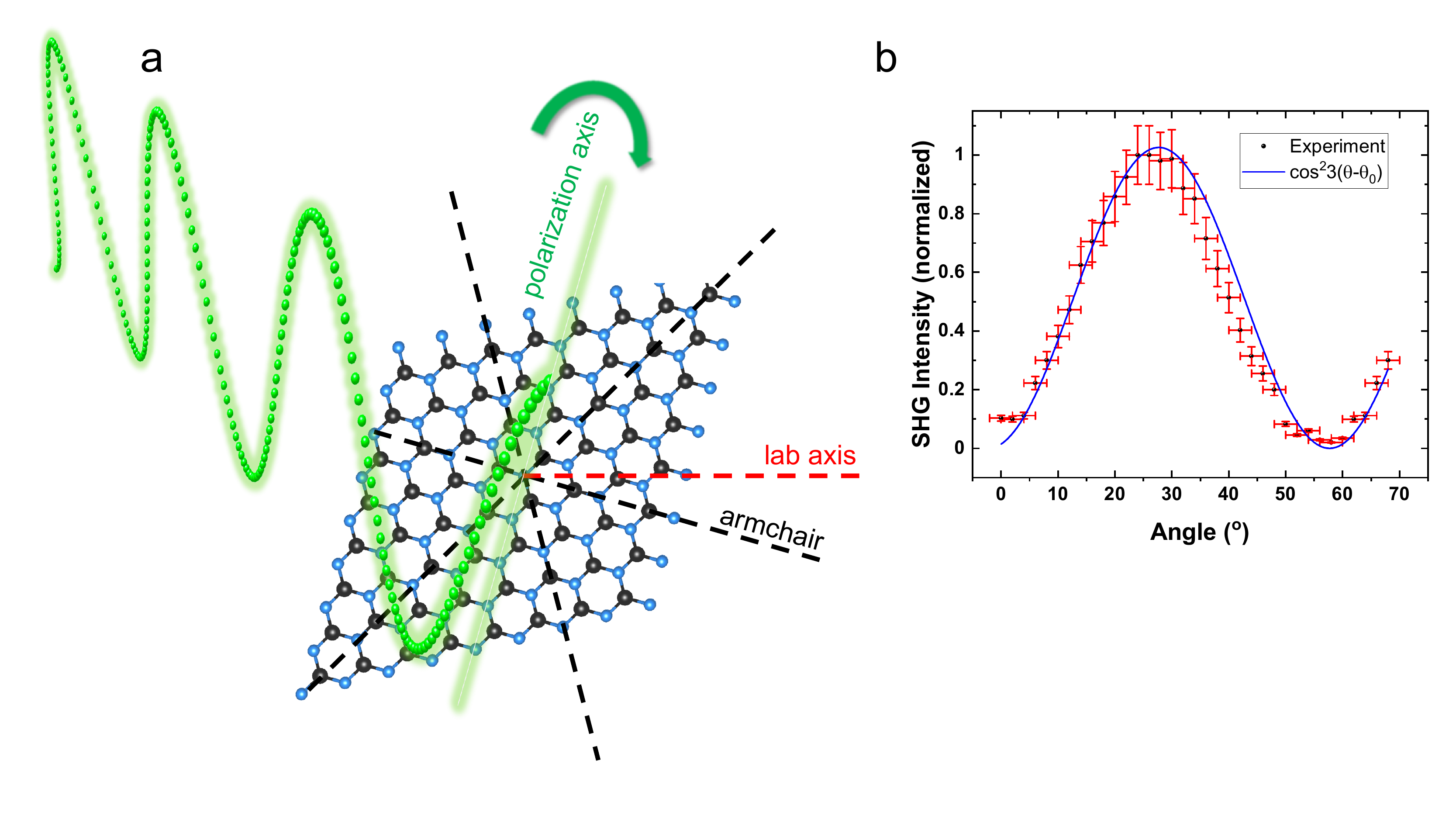}}
\caption{\label{fig:figS3} (a) Schematic representation of a monolayer TMD structure with respect to the lab axis and the polarization axis of the excitation laser. (b) Example of a PSHG plot where the intensity error and the angle error are considered for the fitting analysis.}
\end{figure*}

In a polarization-resolved SHG experiment, we fix the excitation energy in the excitonic state of interest. Using a superachromatic half-wave plate before the sample (see Fig.~\ref{fig:figS1}b), we rotate the polarization axis of the excitation field (green line in Fig.~\ref{fig:figS3}a) to scan the sample in different angles. A TMD monolayer belongs to the $ D_{3h} $ space group, with non-vanishing second order susceptibility elements along the armchair orientation of the sample (black, dashed lines in Fig.~\ref{fig:figS3}a). The angle of the armchair orientation of the different samples is extracted with respect to the lab axis (red, dashed line in Fig.~\ref{fig:figS3}a), here set at zero degrees. Setting the reference point of the lab axis, we can extract the relative crystallographic orientation between different samples according to the angle-dependent modulation of the SHG intensity.\\
\indent To extract the error bar in the twist angle, different sources of error are considered. A twist angle is calculated by the subtraction between the armchair orientations of two separate PSHG data. Thus, the final error is simply the error propagation in this subtraction. Consequently, it is necessary to approximate the error of the armchair orientation of each PSHG measurement. For this, we take into account contributions from both the error in the angle of the polarization axis and the intensity. In a typical PSHG plot, the Y values correspond to the intensity of the SHG and the X values to the polarization angle of the excitation laser (Fig.~\ref{fig:figS3}b). Keep in mind that fluctuations in the SHG intensity can also affect the precision on the determination of the armchair orientation. For the intensity fluctuations we take as error $\pm$10$ \% $ of the measured SHG intensity. This is a reasonable error, extracted by real-time monitoring of the SHG intensity. For the X values, the main error source is the step size of the half-wave plate. In our PSHG experiments we turn the half-wave plate 1$ ^{o} $ before each measurement. This corresponds to 2$ ^{o} $ shift in the polarization angle of the laser. As a result, the error here is $ \pm $ 2$ ^{o} $. The X and Y error values are taken into account in the weighting method during the cosine function fit (Fig.~\ref{fig:figS3}b). The weights will be used in the procedure of reducing the chi-square of the fit, while the weight formula used here is $ W_{i}=\dfrac{1}{\sigma^{2}_{i}}$, with $\sigma_{i}$ the error bar size of each point.

\subsection{D. Twist angle comparison between \textit{in-situ} experiments and on bare monolayers}

\begin{figure*}
\centerline{\includegraphics[width=140mm]{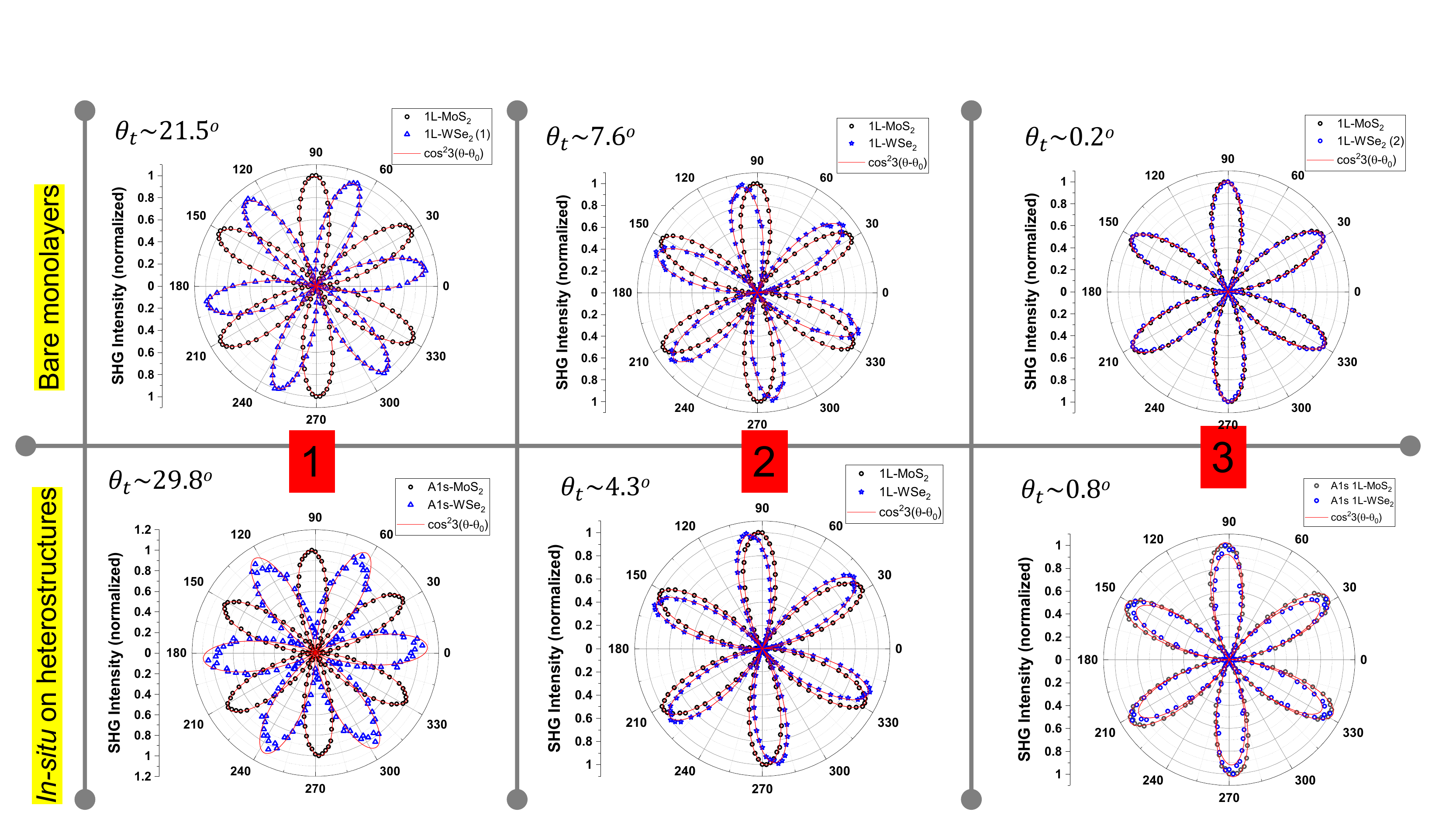}}
\caption{\label{fig:figS4} Comparison between polar plots. Top row shows PSHG plots extracted on the bare monolayers, outside the heterostructure. Bottom row shows PSHG plots extracted directly on the heterostructure by selecting twice the excitation energy to be resonant with the A states. The extracted values of the twist angle, $ \theta_{t} $, are presented in each case. Black and blue points correspond to MoS$_2$ and WSe$_2$, respectively.}
\end{figure*}

Comparing the extracted twist angle when the excitation laser spot is on the bare monolayers or directly on the heterostructure reveals a good agreement. In Fig.~\ref{fig:figS4} we present PSHG data when the excitation spot was on the bare monolayers (top row) and directly on the heterostructure (bottom row). In the latter case, we tune twice the excitation energy on the A-excitonic resonance of MoS$_2$ (black points) and WSe2$_2$ (blue points). As a result, the contribution of the total SHG signal originates predominantly from one of the two monolayers, allowing access to the twist angle directly on the heterostructure. The extracted twist angles, $ \theta_{t} $, between the two methods are similar. However, in some cases there is a small angle difference. For instance in HS1 we get $ \theta_{t} $=29.8$ ^{o} $ with the $in-situ$ approach, whereas $ \theta_{t} $=21.5$ ^{o} $ when we measure on the bare monolayers, outside the heterostructure. Deviations between the two type of measurements can be possibly attributed to local variations of the crystal structure, such as strain and/or disorder caused by impurities/contaminants, etc. In Fig.~\ref{fig:figS4} we note that the MoS$_2$ monolayer orientation is the same in all panels, as can be expected. The twist angle difference for HS1 between the monolayer part inside compared to outside the stack could originate from a different orientation of the WSe$_2$ after full transfer.


\end{document}